\documentclass[%
 reprint,
 amsmath,amssymb,
 aps,
prd,
floatfix,
]{revtex4-2}

\usepackage[pdftex]{graphicx}

\usepackage{float}
\usepackage{amsmath}
\usepackage{dcolumn}
\usepackage{bm}
\usepackage{multirow}
\usepackage{xcolor}
\usepackage[colorlinks=true, linkcolor=blue, citecolor=blue, urlcolor=blue, breaklinks=true]{hyperref}
\usepackage{booktabs}
\usepackage{here}
\newcommand{\mnras}{Monthly Notices of the Royal Astronomical Society}
\newcommand{\mycomments}[1]{#1}
\newcommand{\mycommentsII}[1]{#1}



\begin{document}


\title{Enhancing the reliability of machine learning for gravitational wave\\
parameter estimation with attention-based models}

\author{Hibiki Iwanaga\textsuperscript{1}}
\email{iwanagahibiki7382@gmail.com}
\author{Mahoro Matsuyama\textsuperscript{1}}
\author{Yousuke Itoh\textsuperscript{1,2}}

\affiliation{\textsuperscript{1}Department of Physics, Osaka Metropolitan University,
Osaka 558-8585, Japan}
\affiliation{\textsuperscript{2}Nambu Yoichiro Institute of Theoretical and Experimental Physics (NITEP),
Osaka Metropolitan University, Osaka 558-8585, Japan}

\date{\today}

\begin{abstract}
We introduce a technique to enhance the reliability of gravitational wave parameter estimation results produced by machine learning. We develop two independent machine learning models based on the Vision Transformer to estimate effective spin and chirp mass from spectrograms of gravitational wave signals from binary black hole mergers. To enhance the reliability of these models, we utilize attention maps to visualize the areas our models focus on when making predictions. This approach enables us to demonstrate that both models perform parameter estimation based on physically meaningful information. Furthermore, by leveraging these attention maps, we demonstrate a method to quantify the impact of glitches on parameter estimation. We show that as the models focus more on glitches, the parameter estimation results become more strongly biased. This suggests that attention maps could potentially be used to distinguish between cases where the results produced by the machine learning model are reliable and cases where they are not.

\end{abstract}

\maketitle

\section{\label{sec:level1}Introduction}
\vspace{-10pt}
Since the first detection of gravitational waves from a binary black hole merger on September 14, 2015 \cite{gw1}, observations by LIGO \cite{ALIGO} and Virgo \cite{AVIRGO} have continued. Combining the results of the O1, O2, and O3 observation runs, a total of 90 gravitational wave events have been detected \cite{gwtc1, gwtc2, gwtc3}.
With future improvements in the sensitivity of the LIGO and Virgo detectors, and with the participation of detectors like KAGRA \cite{KAGRA} and LIGO-India \cite{LIGO-India}, the number of detected events is expected to continue increasing \cite{Prospects}.

In gravitational wave data analysis, when the gravitational wave source is a binary merger event, once a detection is made, a parameter estimation based on Bayesian inference is performed \cite{AbbottNoise}. To estimate the posterior distribution, parameter space sampling is typically performed using stochastic sampling methods such as MCMC \cite{emcee} or Nested Sampling \cite{dynesty}. Tools such as LALInference \cite{LALInference} and Bilby \cite{BILBY} have been developed to implement these algorithms.
However, these methods face the challenge of significant computational costs, primarily due to the time required for waveform generation and likelihood function evaluation. Depending on the signal duration and the complexity of the waveform model, it is known that the total time required for estimation can extend to several days per event for binary black hole mergers \cite{gwtc1} and even several weeks for binary neutron star mergers \cite{gw170817}. This problem is expected to become more serious as the number of detected gravitational wave events increases exponentially in the future \cite{Future_detection}.

Several efforts have recently been made to reduce the computational cost of gravitational wave parameter estimation using machine learning.
In \cite{2022arXiv220111126M}, the authors achieved fast analysis by using a CNN to estimate the mass parameters of binary black hole mergers. Additionally, for the events in GWTC-1, they confirmed that the results were consistent with the posterior distributions obtained from LALInference.

In \cite{ChuaAlvinJ} and \cite{Gabbard2022}, Bayes' rule is applied to enable network training using only samples from the prior distribution and gravitational wave likelihood. This approach allows for the estimation of posterior distributions without the need for posterior samples during training. 
As noted above, likelihood evaluation requires significant computational costs. Therefore, these so-called ``likelihood-free" methods are highly valuable, leading to the exploration of many potential applications \cite{Green2020,2020PhRvD.102j4057G,Shen2022}.
A notable example is the DINGO framework \cite{DINGO}, which uses normalizing flows to estimate all 15 parameters of BBH mergers, achieving results consistent with the posterior distributions obtained from LALInference. In the DINGO framework, the neural network is conditioned not only on the strain data but also on the power spectral density (PSD) of the detector noise, effectively addressing the fact that detector noise varies from event to event. This excellent framework has also been expanded to parameter estimation for binary neutron stars \cite{DINGO-BNS}.
Therefore, machine learning is progressing toward practical applications not only in gravitational wave detection and glitch classification but also in parameter estimation (See, e.g., \cite{2021arXiv211106987C,2021MLS&T...2a1002C,2024AnP...53600140S,2022NatSR..12.9935S}). (For typical approaches that do not use machine learning, such as reduced-order modeling, efficient sampling methods, and others, see, e.g., \cite{2021PhRvD.104d4062M,2022PhRvD.105l4057L,2022PhRvD.106j4053A,2023PhRvD.108l3040M} and references therein.)

When applying machine learning to gravitational wave parameter estimation, it is important to consider the ``black-box" problem \cite{Ribeiro,Mythos}.
Because machine learning models learn from large amounts of data and autonomously generate outputs, their decision-making process is unclear. 
For example, even if there are significant noises such as glitches nearby gravitational wave signals, machine learning models generate outputs without warning. For another example, it has not been sufficiently verified whether the models are focusing on physically meaningful information.

\mycomments{In this work, we used Data-efficient Image Transformers (DeiT) \cite{deit}, which is a model based on ViT \cite{ViT}. 
ViT applies the transformer architecture\cite{Transformer}, originally designed for natural language processing, to image data.
ViT utilizes image patch embedding, positional embeddings, transformer encoder, and classification token. While ViT achieved the state-of-the-art results in image classification when it appeared,  it requires huge training datasets and it is computationally expensive for small datasets. DeiT has then successfully reduced the computational cost of training while maintaining accuracy by utilizing knowledge distillation \cite{knowledge}  and data augmentation.  We adopted DeiT due to our limited availability of computing resources.  We make use of attention maps produced by DeiT that show the areas a machine learning model focuses on when it makes predictions, thereby improving the reliability of outputs. }
\mycommentsII{See the appendix \ref{sec:appendix_Deit} for further details about the Deit network we used.}

\vspace{-10pt}
\section{Method}
\vspace{-10pt}
Our main purpose is to verify whether the machine learning models focus on physically meaningful information through the Attention Maps. Therefore, we used the spectrogram as it is intuitive to understand, making it easier to review and discuss the results afterward.
As a first step, we build models to estimate the effective spin and chirp mass of binary black hole mergers. To examine the differences in attention regions for the estimation of each parameter, we prepared two independent models.
In this section, we describe the details of the dataset used for training our models.

\vspace{-10pt}
\subsection{GW dataset generation}\label{subsection:data_generation}
\vspace{-10pt}
For the generation of CBC waveforms, we used PyCBC \cite{pycbc} with the SEOBNRv4\_opt \cite{SEOBNRv4_opt} waveform model.
Our models were trained with parameters from Table~\ref{tab:param_range}. Although some parameters are restricted, we chose a simple configuration for the sake of simplicity, considering our main purpose.

Regarding the data generation procedure, we first created strain data by injecting a waveform into 4 second long data of stationary Gaussian noise colored by the LIGO design sensitivity (aLIGOZeroDetHighPower) \cite{pycbc}. During this process, the central time of the noise data was aligned with the merger time of the waveform. \mycomments{Then, we whitened the strain time series and cut the first and last 0.5 seconds of data to minimize possible edge effects. 
After the whitening process, we applied the constant-Q transform to create the spectrogram \cite{2004CQGra..21S1809C}. The constant-Q transform is a type of time-frequency transform similar to the Fourier transform. The constant-Q transform has high frequency resolution at low frequencies and high time resolution at high frequencies. Depending on the properties of data, the constant-Q transform is  
beneficial for catching detailed spectral structures than the usual Fourier transform.}

This spectrogram covers a frequency range of $f_{\rm min}=20$ [Hz] to $f_{\rm max}=1024$ [Hz] and includes 3 seconds of data.
Following this process, we generated 10,000 spectrograms and split them into training and validation sets in a 9:1 ratio to train the models.

\begin{table}[H]
\centering
    \begin{tabular}{@{}l>{\centering\arraybackslash}m{2cm}>{\centering\arraybackslash}m{2cm}@{}}
        \toprule
        & \textbf{Range} \\
        \midrule
        Source-frame chirp mass $\mathcal{M}/M_\odot$ &  [20, 40]  \\
        Mass ratio $q$ &  [0.4, 1.0]  \\
        Effective spin $\chi_{\text{eff}}$ &  [-0.88, 0.88]  \\
        Luminosity distance $d_L$/Mpc &  [50, 600]  \\
        Inclination angle $\theta_{JN}$ &  [0, $\pi$]  \\
        Polarization angle $\psi$ &  [0, $\pi$]  \\
        \bottomrule
    \end{tabular}
    \caption{Parameters used to create the training and validation datasets. $\mathcal{M}$, $q$, $\chi_{\text{eff}}$, $\theta_{JN}$, and $\psi$ are uniformly distributed. Source-frame component masses $m_1$,$m_2$, are determined from $\mathcal{M}$ and $q$, while the dimensionless projections of the individual BH spins $\chi_1$ and $\chi_2$ are determined from $\chi_{\text{eff}}$. 
The luminosity distance $d_L$ and the sky position (right ascension $\alpha$ and declination $\delta$) are set to ensure a uniform distribution of gravitational wave sources within a sphere.}
    \label{tab:param_range}
\end{table}

\subsection{Training and validation}
\mycomments{As was mentioned in the introduction, we used Data-efficient Image Transformers (DeiT) \cite{deit} that is pre-trained on ImageNet \cite{ImageNet} and we trained it on the dataset we created.}

\begin{figure}[H]
    \centering
    \includegraphics[width=0.8\linewidth]{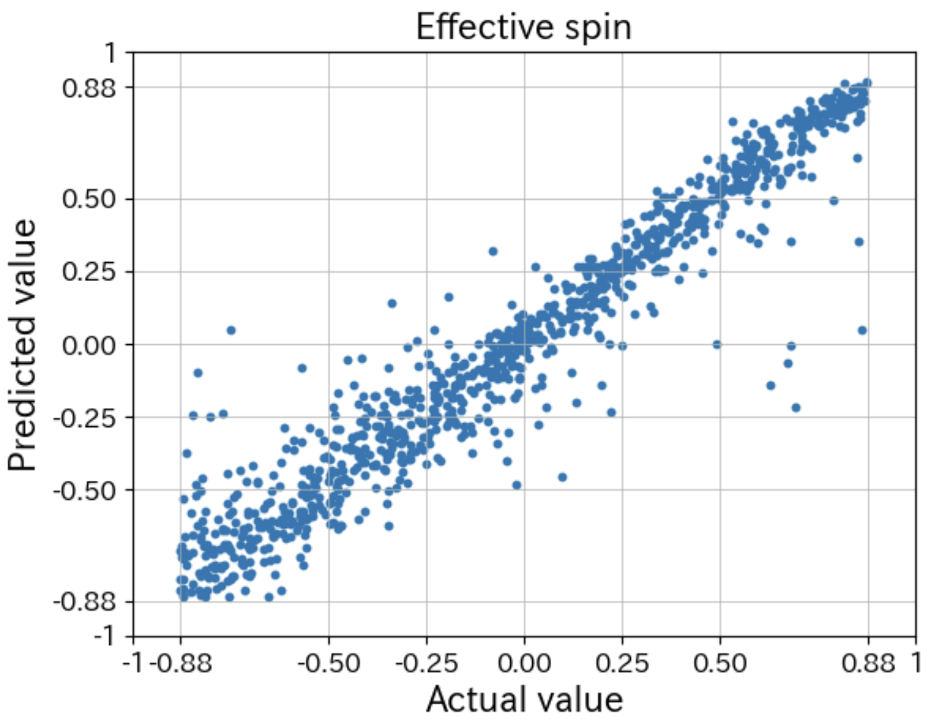}
    \caption{A predicted vs actual plot after 50 epochs for the effective spin estimation model. The model was trained for 50 epochs with a learning late of $1 \times 10^{-4}$ and a batch size of 16.}
    \label{fig:scatter_eff_spin}
\end{figure}

\begin{figure}[H]
    \centering
    \includegraphics[width=0.8\linewidth]{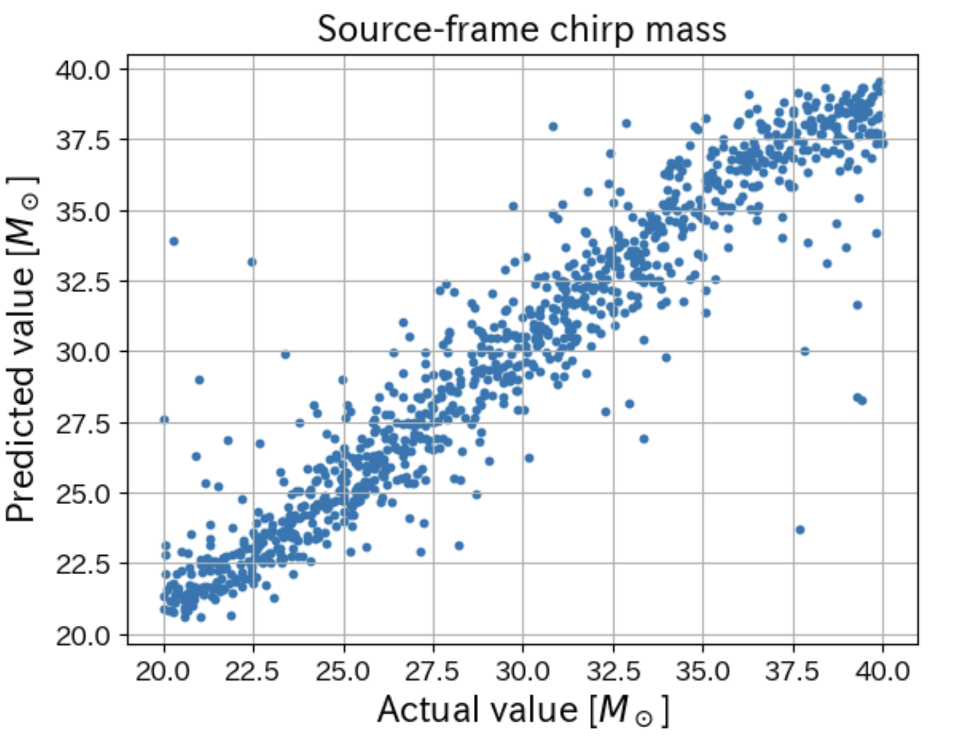}
    \caption{A predicted vs actual plot after 50 epochs for the chirp mass estimation model. The model was trained for 50 epochs with a learning late of $5 \times 10^{-5}$ and a batch size of 16.}
    \label{fig:scatter_chirp}
\end{figure}

Our models were trained for 50 epochs with a batch size of 16. The Adam optimizer \cite{Adam} was used for optimization. Fig.~\ref{fig:scatter_eff_spin} and~\ref{fig:scatter_chirp} presents scatter plots illustrating the relationship between predicted and actual values for the validation dataset, after 50 epochs of training and validation. 

Based on these results, it was confirmed that our models successfully learned the characteristics of the effective spin and chirp mass. The training took approximately one hour on an NVIDIA GeForce RTX 3090.

\vspace{1em}

\vspace{-10pt}
\subsection{Evaluation of uncertainty}\label{subsection:uncertainty}
\begin{figure}[H]
    \centering
    \includegraphics[width=0.75\linewidth]{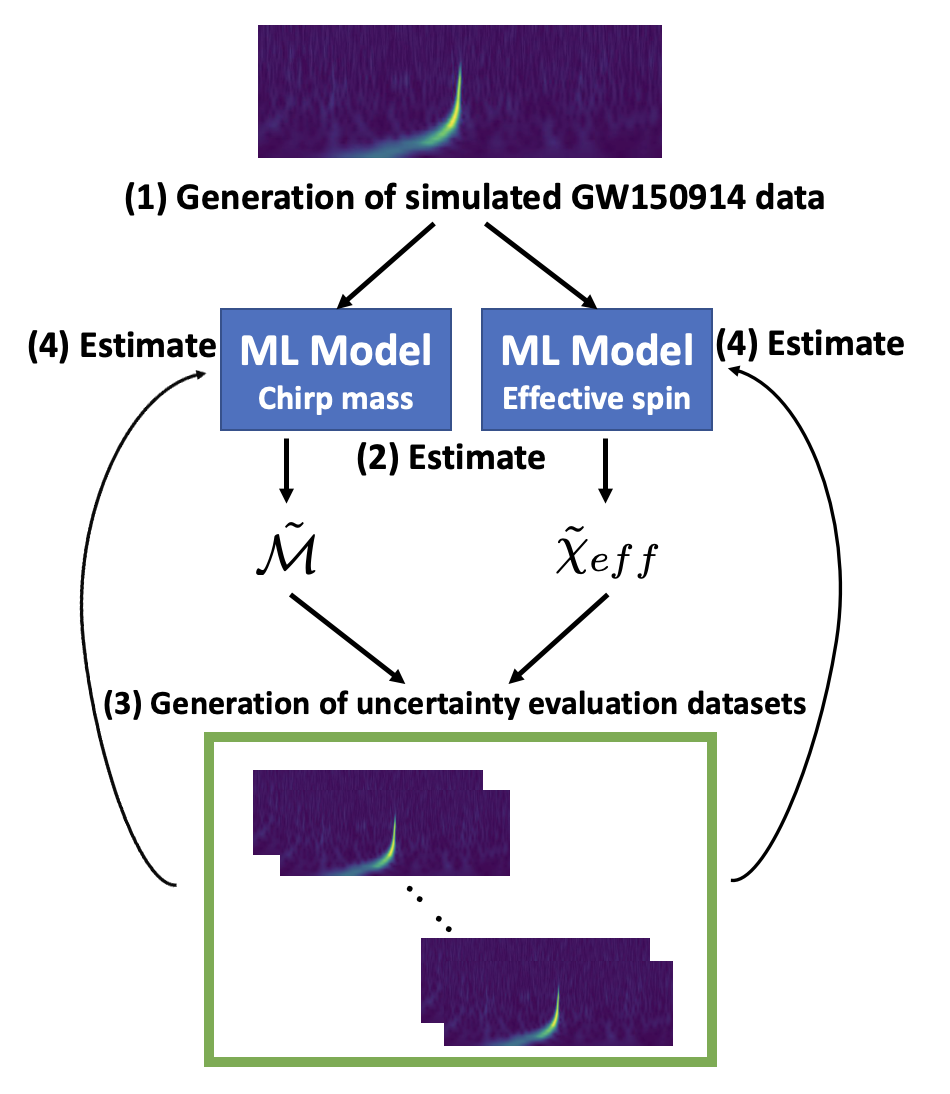}
    \caption{Flow chart of uncertainty evaluation. 
    }
    \label{fig:flowchart}
\end{figure}

\vspace{-10pt}
To evaluate the uncertainty in the predictions of our models, we developed an uncertainty evaluation method based on a primitive Monte Carlo approach. \mycomments{See the appendix \ref{sec:appendix} for further explanation and possible use cases of our method.}
The flowchart of this method is shown in Fig.~\ref{fig:flowchart}.

(1) Initially, a waveform was generated using parameters consistent with the estimates obtained from LALInference for GW150914. After injecting this waveform into stationary Gaussian noises colored by the same PSD as in Sec.~\ref{subsection:data_generation}, a constant-Q transform was applied to create a spectrogram. 

(2) Using this spectrogram as input to our models, we estimated the effective spin and chirp mass.

(3) Subsequently, a waveform was generated based on the values estimated in (2). For parameters that were not estimated, the values were set to those used in (1). This waveform was injected into stationary Gaussian noises colored by the same PSD as in Sec.~\ref{subsection:data_generation}, and a constant-Q transform was applied to create spectrograms. In this way, we created an uncertainty evaluation dataset that contains 1,000 spectrograms.

(4) We evaluated absolute errors by using our models to make predictions on this dataset. Then we calculated 90\% confidence intervals for our models.

\vspace{-20pt}

\section{Results}

\vspace{-5pt}
\subsection{Prediction results}\label{subsection:prediction}
\begin{figure}[H]
    \centering
    \includegraphics[width=0.8\linewidth]{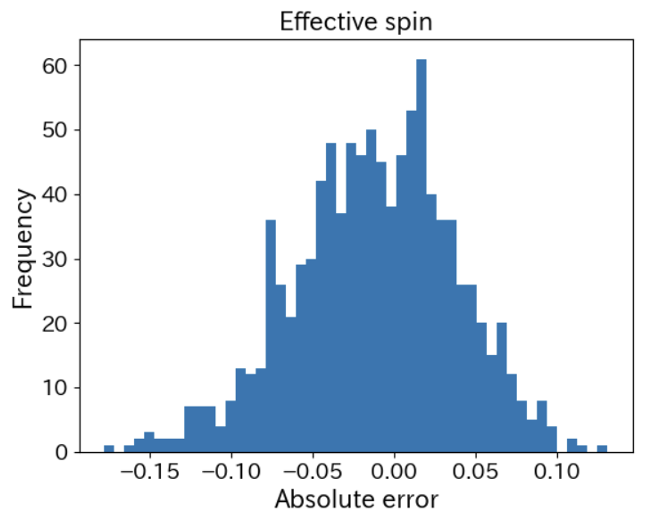}
    \caption{Histogram of absolute error for the effective spin estimation model. The absolute errors are calculated from the effective spins estimated in Sec.~\ref{subsection:uncertainty}-(4) and the estimated value in Sec.~\ref{subsection:uncertainty}-(2).}
    \label{fig:abs_error_spin}
\end{figure}

\begin{figure}[H]
    \centering
    \includegraphics[width=0.8\linewidth]{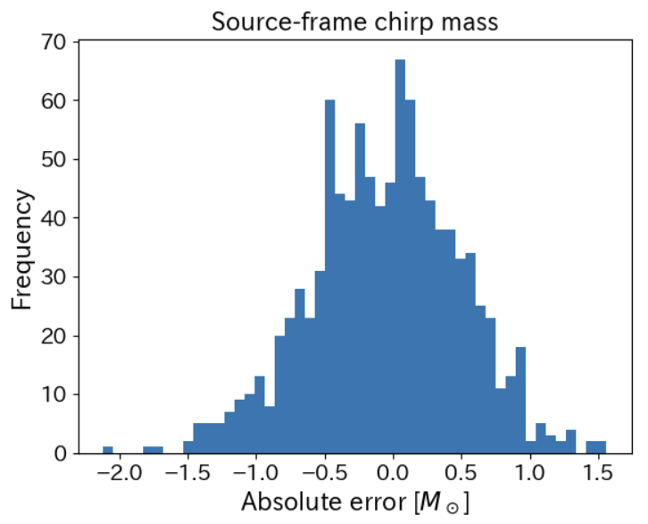}
    \caption{Histogram of absolute error for the chirp mass spin estimation model. The absolute errors are calculated from the chirp masses estimated in Sec.~\ref{subsection:uncertainty}-(4) and the estimated value in Sec.~\ref{subsection:uncertainty}-(2).}
    \label{fig:abs_error_chirp}
\end{figure}

\vspace{-10pt}
Table~\ref{tab:comparison} shows the 90\% confidence intervals, while Fig.~\ref{fig:abs_error_spin} and~\ref{fig:abs_error_chirp} illustrate the distributions of absolute errors. In Table~\ref{tab:comparison}, it can be seen that the predictions of our models were approximately consistent with those obtained by LALInference for GW150914. However, we did not analyze real GW events, and since the PSD used corresponds to the design sensitivity, this represents the result of a rough comparison. \mycommentsII{We also remark that since we used a frequentist approach while the results by LALInference are based on a Bayesian approach, two results are not necessarily expected to be the same.}

The total estimation time was approximately six minutes, confirming that computational costs can be significantly reduced. This method does not require extensive waveform generation and likelihood evaluations, making it feasible to achieve a similar order of efficiency when applied to real GW data.

\begin{table}[H]
\centering
\begin{tabular}{@{}l>{\centering\arraybackslash}m{3cm}>{\centering\arraybackslash}m{3cm}@{}}
\toprule
  & Our models & LALInference\\ \midrule
$\chi_{\text{eff}}$ & $-0.09^{+0.07}_{-0.09}$ & $-0.09^{+0.19}_{-0.17}$ \\
[1.2ex]
$\mathcal{M}/M_\odot$\fontsize{8}{8}\selectfont & $27.6^{+0.8}_{-1.0}$ & $27.9^{+2.3}_{-1.8}$ \\ \bottomrule
\end{tabular}
    \caption{Comparisons of predictions between our models and LALInference. Our results provide the 90\% confidence intervals, while the LALInference results are the 90\% credible intervals estimated using the EOBNR waveform, as reported in \cite{GW150914properties}. Note that we used stationary Gaussian noise and assuming design sensitivity.}
    \label{tab:comparison}
\end{table}

\vspace{-15pt}

\vspace{-15pt}

\subsection{Attention Map}
\begin{figure}[H]
    \renewcommand{\thefigure}{6}
    \centering
    \includegraphics[width=\linewidth]{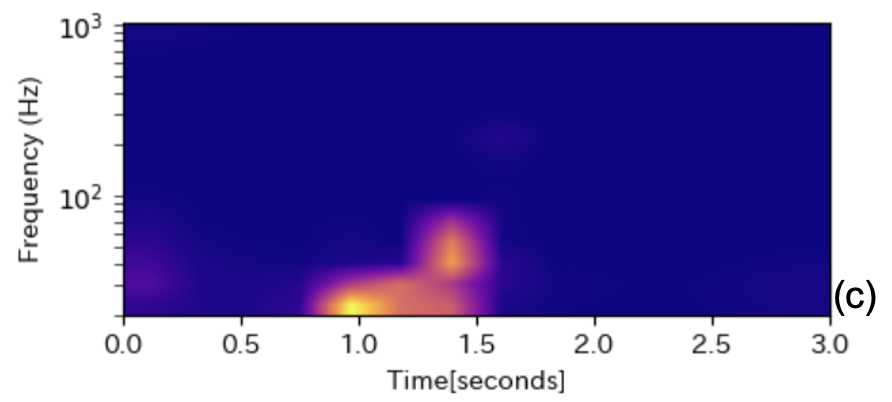}
    \caption{Attention maps for the uncertainty evaluation dataset of simulated GW150914. (a) An example of input data. (b) Composite of 1000 overlaid attention maps for effective spin estimation. (c) Composite for chirp mass. All attention maps are composed of 14×14 pixels and are resized to match the 224×224 resolution of the input data using bilinear interpolation.}
    \label{fig:Attention_map_gw150914_chirp}
\end{figure}

\vspace{-10pt}
We generated attention maps for the 1,000 spectrograms used in the uncertainty evaluation described in Sec.~\ref{subsection:uncertainty}. 
\mycomments{Each attention map we used is for the class token at the last self-attention layer of the network.} 
To illustrate the overall trend, we created a composite figure by overlaying these attention maps (Fig.~\ref{fig:Attention_map_gw150914_chirp}). It can be seen that the effective spin estimation model appears to focus primarily on the mid-frequency range, whereas the chirp mass estimation model focuses on the low-frequency range. 

Each intrinsic binary parameter has a specific frequency range where its information predominantly comes from.
It is known that the chirp mass tends to appear more prominently at lower frequencies, while the effective spin shows its effect at mid-to-high frequencies \cite{blanch,fisher}.
From the attention map results, we confirmed that our models focus on the ranges where the effects of each parameter are predominant.
These results suggest that the predictions of our models are based on physically meaningful information.

\begin{figure}[H]
    \renewcommand{\thefigure}{6.1}
    \centering
    \includegraphics[width=\linewidth]{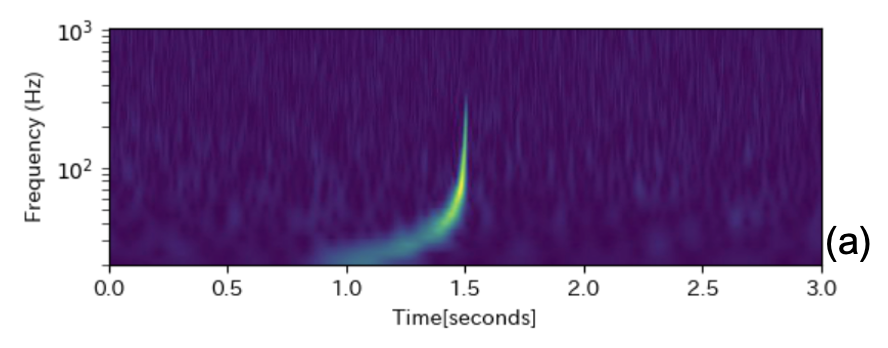}
    \label{fig:Input_GW150914}
\end{figure}

\vspace{-20pt}

\begin{figure}[H]
    \renewcommand{\thefigure}{6.2}
    \centering
    \includegraphics[width=\linewidth]{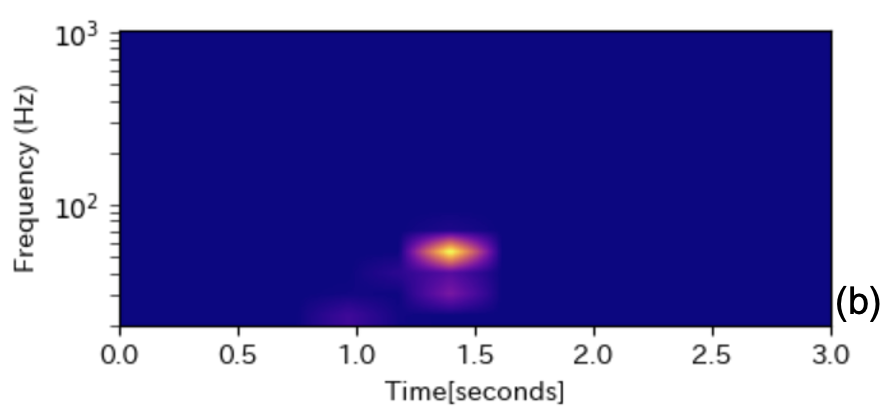}
    \label{fig:Attention_map_gw150914_spin}
\end{figure}

\vspace{-20pt}

\renewcommand{\thefigure}{\arabic{figure}}

\vspace{-10pt}

\subsection{Attention Maps for a longer GW signal}
\vspace{-10pt}
To more clearly see the differences in the attention regions of the two models, we conducted the same analysis using a waveform characterized by a large effective spin and a small chirp mass, which results in a relatively longer signal duration.

\begin{figure}[H]
    \renewcommand{\thefigure}{7}
    \centering
    \includegraphics[width=\linewidth]{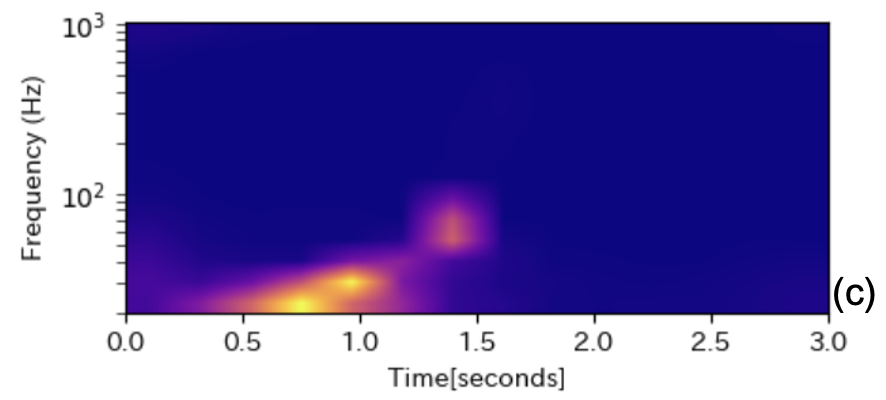}
    \caption{Attention maps for the uncertainty evaluation dataset of simulated longer GW event. (a) An example of input data. (b) composite of 1000 overlaid attention maps for effective spin estimation. (c) Composite for chirp mass.  All attention maps are composed of 14×14 pixels and are resized to match the 224×224 resolution of the input data using bilinear interpolation. }
    \label{fig:Attention_map_longerGW_chirp}
\end{figure}

We generated a waveform with an effective spin of 0.8 and a source-frame chirp mass of 22$M_\odot$, and performed uncertainty evaluation following the same procedure as in section ~\ref{subsection:uncertainty}. We then created a composite figure by overlaying 1000 attention maps (Fig.~\ref{fig:Attention_map_longerGW_chirp}). It can be seen that the two models are focusing on the mid-frequency range and low-frequency range, respectively.

\begin{figure}[H]
    \renewcommand{\thefigure}{7.1}
    \centering
    \includegraphics[width=\linewidth]{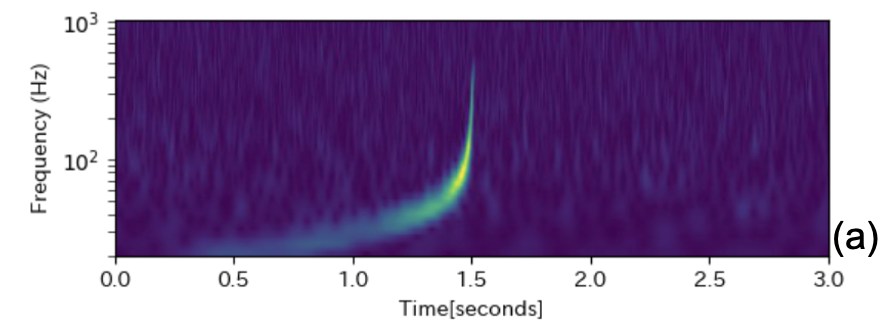}
    \label{fig:Input_longerGW}
\end{figure}

\vspace{-20pt}

\begin{figure}[H]
    \renewcommand{\thefigure}{7.2}
    \centering
    \includegraphics[width=\linewidth]{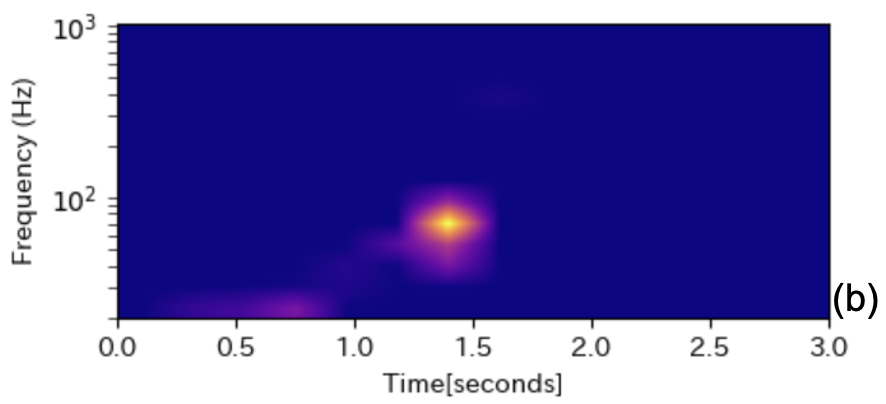}
    \label{fig:Attention_map_longerGW_spin}
\end{figure}

\vspace{-20pt}

\renewcommand{\thefigure}{\arabic{figure}}

\section{Application of Attention Maps for Evaluating the Impact of Glitches}
\vspace{-10pt}
In real strain data, transient noise artifacts, commonly known as glitches, occasionally occur nearby gravitational wave signals \cite{LIGOdetchar,VIRGOdetchar,KAGRAdetchar}. These glitches are problematic as they not only generate false-positive candidates \cite{FP1,FP2} but also bias parameter estimation \cite{PEbias1,PEbias2,PEbias3,PE_bias_gw191109}.
Therefore, to apply machine learning to the parameter estimation of gravitational waves, it is essential to properly evaluate the impact of glitches and enhance the reliability of the results.

In this study, we demonstrate the potential of utilizing attention maps to quantify the impact of glitches on parameter estimation.

\subsection{Method}

\vspace{-10pt}
We injected glitches into the simulated GW150914 data and applied the same uncertainty evaluation method as in Sec.~\ref{subsection:uncertainty} to estimate the extent to which our model's predictions would be biased if glitches were present in the simulated GW event. Furthermore, we examined the relationship between the Attention Map and the prediction results.

For simplicity, we modeled glitches by sine-Gaussian waveforms \cite{sine-gauss}.
The sine-Gaussian waveform can be written in the time domain as
\setlength{\abovedisplayskip}{10pt}
\setlength{\belowdisplayskip}{10pt}
\begin{equation}
h_{\rm sg} = A_{\rm sg} e^{-\frac{(t - t_0)^2}{\mycomments{\tau^2}}} \cos\left(2\pi f_0 (t - t_0)\right)
\end{equation}
$A_{\rm sg}$ is an overall amplitude, $t_0$ and $f_0$ are the location of the sine-Gaussian in time and frequency, respectively, \mycomments{and $\tau$ is the damping time}.

\begin{figure}[H]
    \centering
    \includegraphics[width=\linewidth]{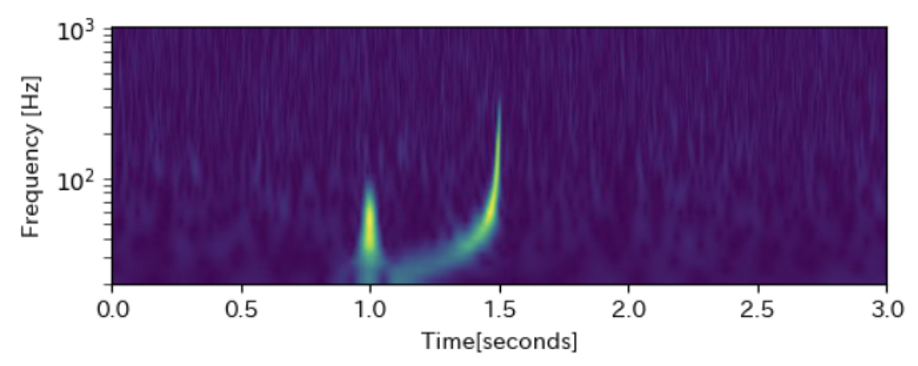}
    \caption{An example spectrogram of simulated GW150914 data with an injected glitch of amplitude $1A$.}
    \label{fig:glitch_1A}
\end{figure}

In real GW data analysis, it is problematic when glitches directly overlap GW signals as in the case of GW170817 \cite{gw170817}.
However, injecting around the merger phase seemed likely to cause significant bias, thereby making it difficult to identify clear trends. Therefore, we selected \mycomments{$\tau=10^{-2}$ [s]}, $f_0 =40$ [Hz] and $t_0=1.0$ [s] to configure the glitches to slightly overlap with the inspiral phase. For the amplitude $A_{\rm sg}$, we prepared three glitches with amplitudes of $0.5A$, $0.8A$, and $1A$, using the maximum amplitude $A$ of the simulated GW signal. In this manner, we generated spectrograms for three simulated GW events with injected glitches. Figure~\ref{fig:glitch_1A} presents an example.

\subsection{Predictions}
\vspace{-10pt}
We compared the prediction results from Sec.~\ref{subsection:prediction} with three simulated GW events with glitches of three different amplitudes.

Table~\ref{tab:comparison_glitch} shows the 90\% confidence intervals, while Fig.~\ref{fig:glitch_prediction_spin} and~\ref{fig:glitch_prediction_chirp} illustrates the distributions of the predicted values. It can be seen that for both effective spin and chirp mass, glitches with larger amplitudes more strongly biased parameter estimation. This tendency is also commonly seen in conventional methods.

\begin{table}[H]
\setlength{\abovecaptionskip}{5pt}
\centering
\begin{tabular}{@{}l@{\hspace{1mm}}>{\centering\arraybackslash}m{1.8cm}@{\hspace{1mm}}>{\centering\arraybackslash}m{1.8cm}@{\hspace{1mm}}>{\centering\arraybackslash}m{1.8cm}@{\hspace{1mm}}>{\centering\arraybackslash}m{1.8cm}@{}}
\toprule
  & \makebox[1.5cm]{No glitches} & 0.5A & 0.8A & 1A \\
\midrule
$\chi_{\text{eff}}$ & $-0.09^{+0.07}_{-0.09}$ & $-0.09^{+0.09}_{-0.07}$ & $-0.07^{+0.08}_{-0.07}$ & $0.04^{+0.07}_{-0.05}$ \\
[1.2ex]
$\mathcal{M}/M_\odot$\fontsize{8}{8}\selectfont & $27.6^{+0.8}_{-1.0}$ & $30.5^{+0.8}_{-1.1}$ & $32.6^{+0.6}_{-1.4}$ & $33.3^{+0.7}_{-1.3}$ \\
\bottomrule
\end{tabular}
\caption{90\% confidence intervals for a simulated GW150914 event without glitches and for three simulated GW events with varying glitch amplitudes.}
\label{tab:comparison_glitch}
\end{table}

\vspace{-10pt}

\begin{figure}[H]
    \setlength{\abovecaptionskip}{3pt}
    \centering
    \includegraphics[width=0.8\linewidth]{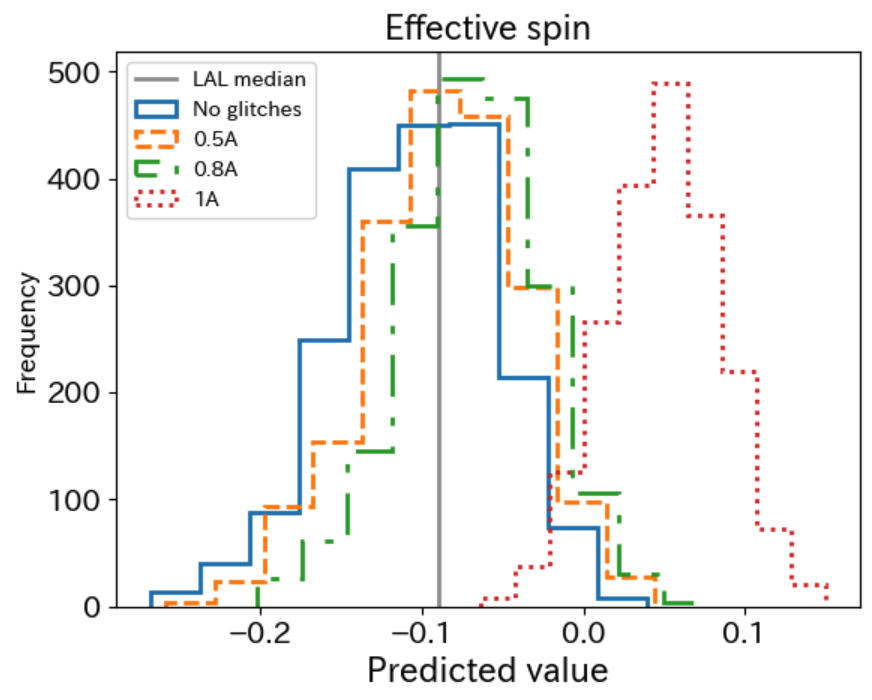}
    \caption{Histograms of the predicted effective spins  for a simulated GW150914 event without glitches and for three simulated GW events with varying glitch amplitudes. The solid vertical lines are the median estimates by LALInference \cite{GW150914properties}.}
    \label{fig:glitch_prediction_spin}
\end{figure}

\vspace{-10pt}

\begin{figure}[H]
    \setlength{\abovecaptionskip}{3pt}
    \centering
    \includegraphics[width=0.8\linewidth]{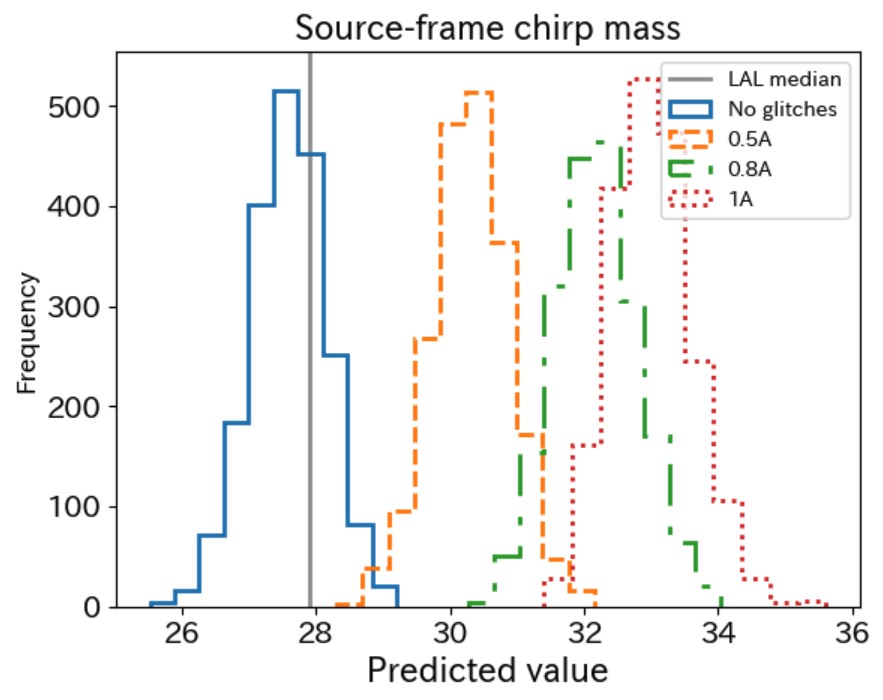}
    \caption{Histograms of the predicted chirp masses for a simulated GW150914 event without glitches and for three simulated GW events with varying glitch amplitudes.}
    \label{fig:glitch_prediction_chirp}
\end{figure}

\subsection{Attention Map}
\vspace{-10pt}
Our goal here is to demonstrate the potential use of Attention Maps for evaluating the impact of glitches on parameter estimation.

\begin{figure*}[!t]
    \centering
    \includegraphics[width=0.8\linewidth]{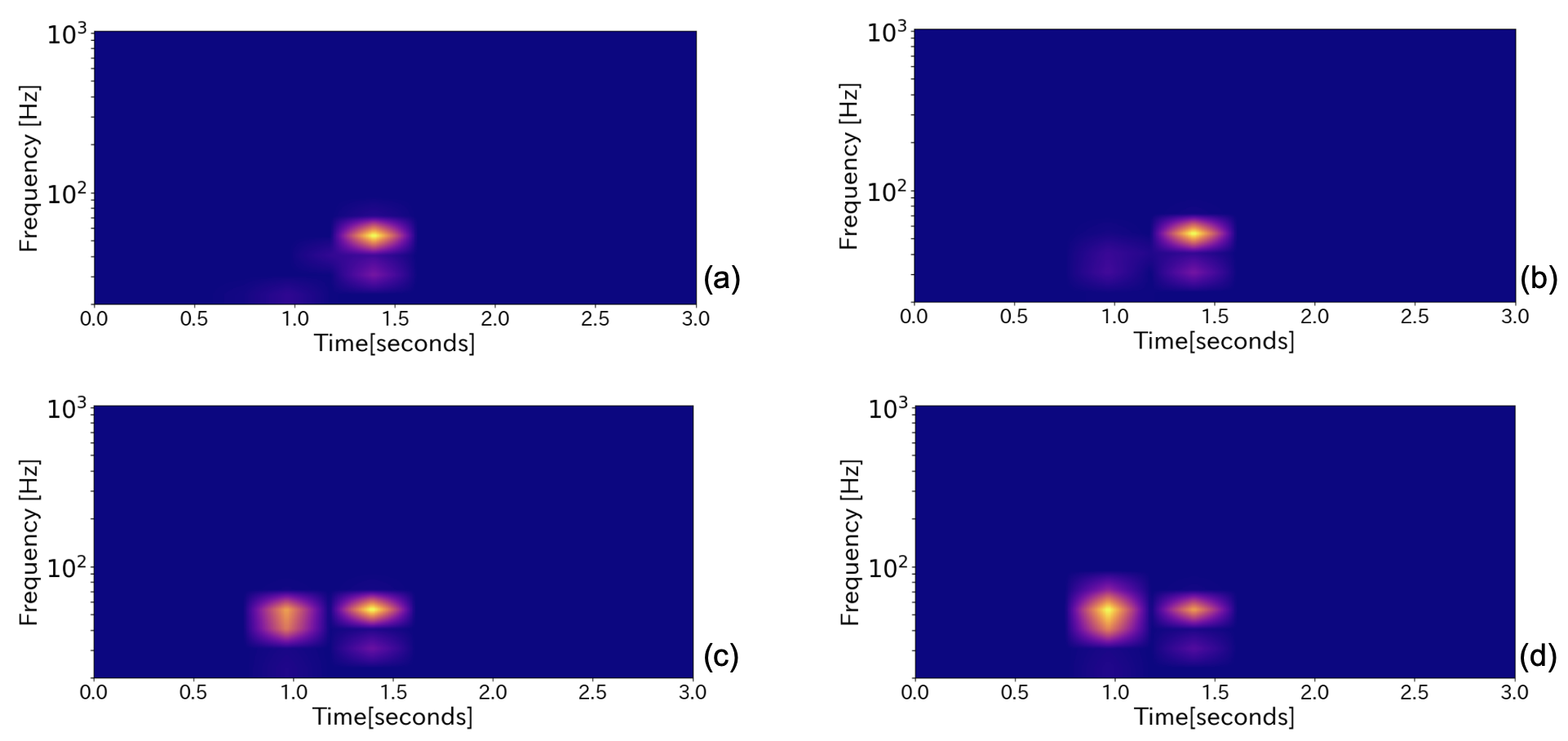}
    \caption{Attention maps for effective spin estimation model. (a) A simulated GW150914 without glitches. Simulated GW events with glitch amplitudes of (b) 0.5A, (c) 0.8A, and (d) 1A.}
    \label{fig:Attention_map_glitch_spin}
\end{figure*}

\begin{figure*}[!t]
    \centering
    \includegraphics[width=0.8\linewidth]{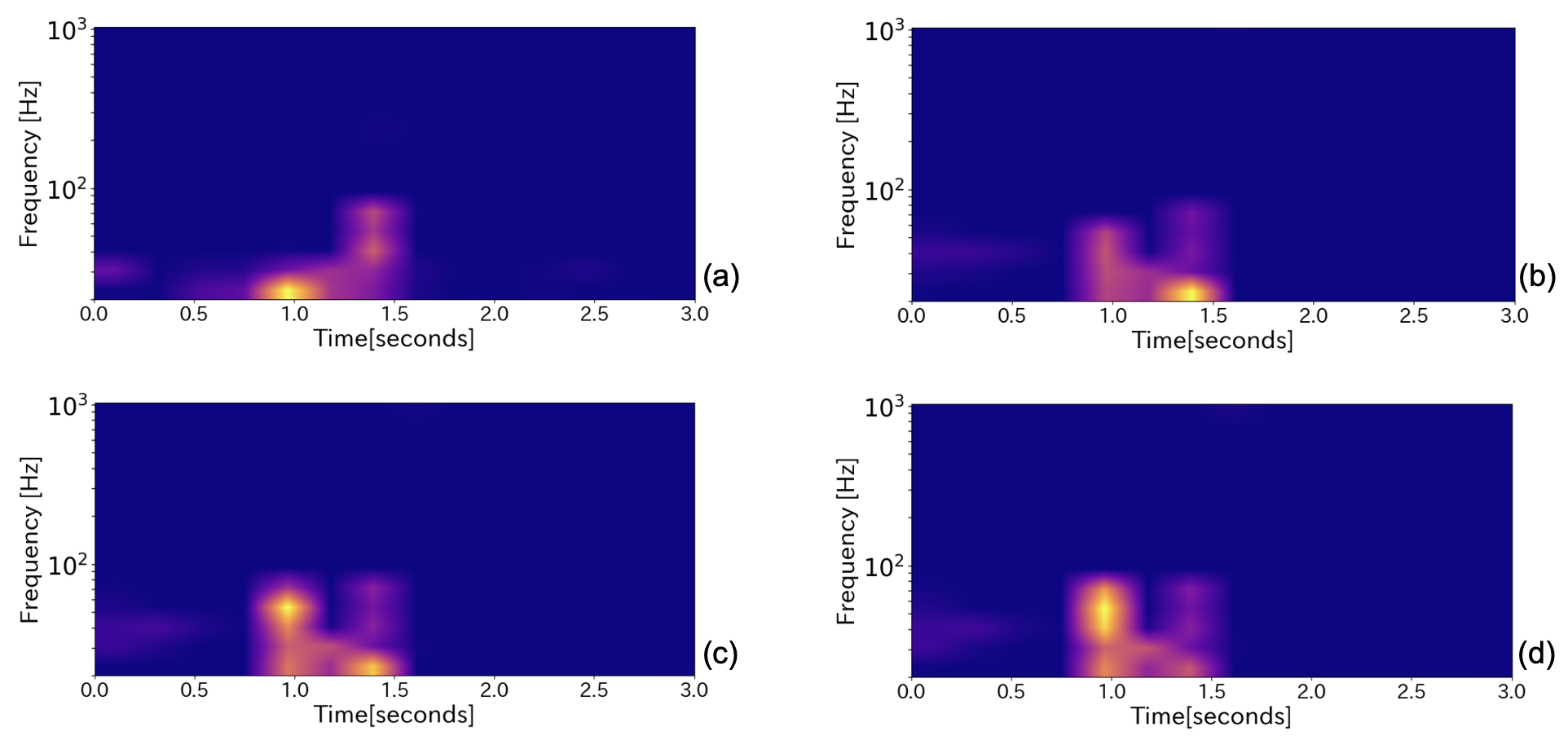}
    \caption{Attention maps for chirp mass estimation model. (a) A simulated GW150914 without glitches. Simulated GW events with glitch amplitudes of (b) 0.5A, (c) 0.8A(c), and (d) 1A.}
    \label{fig:Attention_map_glitch_chirp}
\end{figure*}

Figures~\ref{fig:Attention_map_glitch_spin} and~\ref{fig:Attention_map_glitch_chirp} shows the Attention Maps for a simulated GW150914 event without glitches and for three simulated GW events with varying glitch amplitudes. 
It can be seen that as the amplitude of the glitch increases, the attention around 1.0 seconds and 40 Hz, where the glitch was injected, also becomes larger.

\begin{figure}[H]
    \centering
    \includegraphics[width=0.8\linewidth]{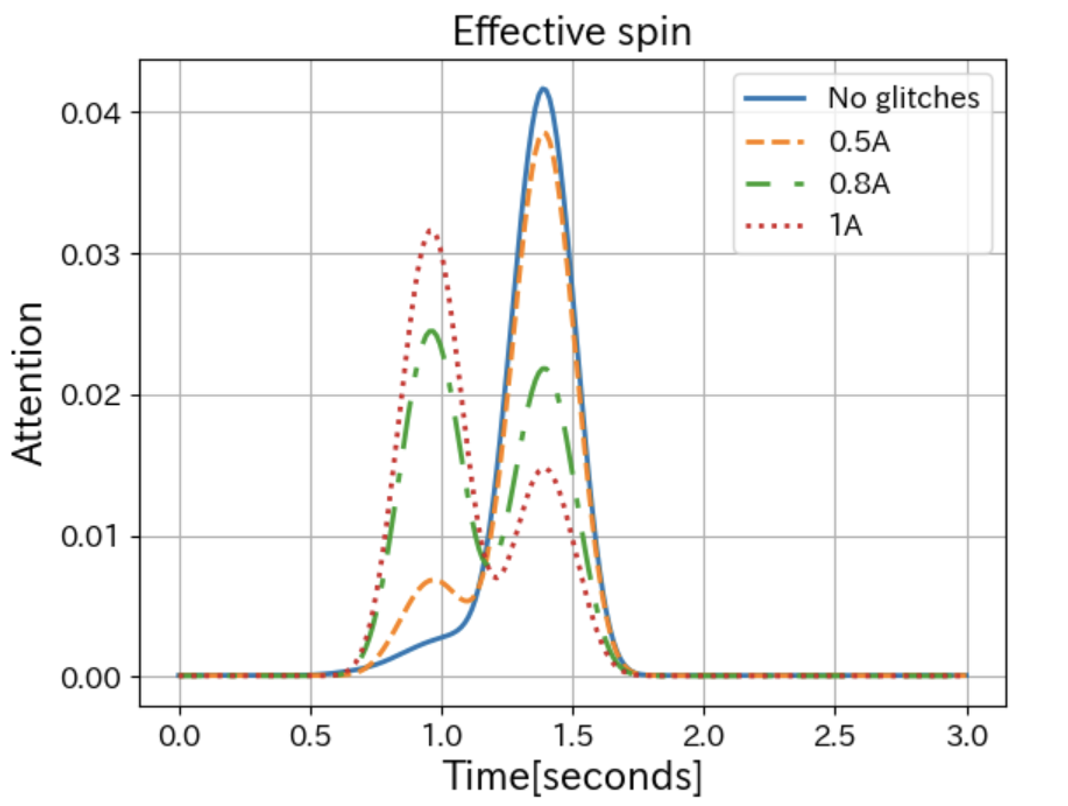}
    \caption{Distributions of the total attention values computed at each time point for effective spin estimation model. All attention values are normalized by scaling between 0 and 1. Additionally, smoothing was performed using a 1D Gaussian filter to account for the resolution of the attention map.}
    \label{fig:Attention_distribution_spin}
\end{figure}

\begin{figure}[H]
    \centering
    \includegraphics[width=0.8\linewidth]{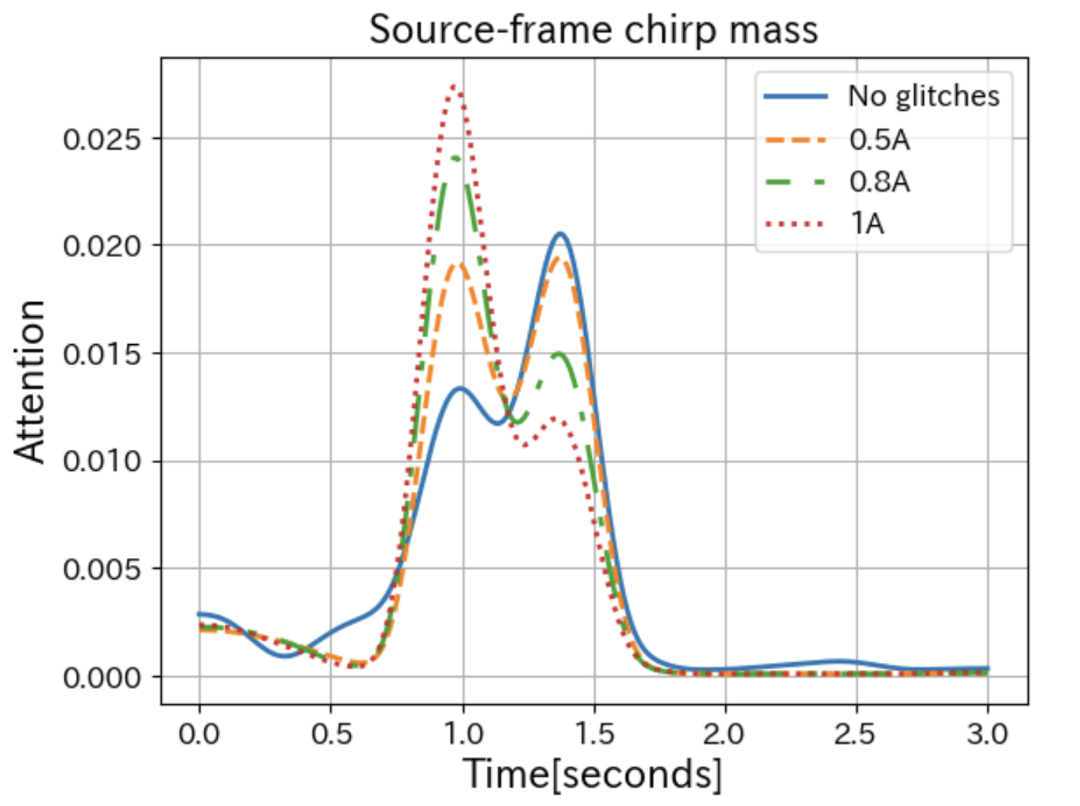}
    \caption{Distributions of the total attention values computed at each time point for chirp mass estimation model.}
    \label{fig:Attention_distribution_chirp}
\end{figure}

Furthermore, to quantify the impact of glitches to some extent, we calculated the total attention values for each time point. The distribution is shown in Figs.~\ref{fig:Attention_distribution_spin} and~\ref{fig:Attention_distribution_chirp}. The figures illustrate that as the amplitude of the glitch increases, the attention values around it also tend to increase.

\vspace{10pt}

\vspace{-10pt}

\newpage

\section{Conclusion}
\vspace{-10pt}
We developed two machine learning models based on the Vision Transformer to estimate effective spin and chirp mass from spectrograms of gravitational wave signals. Predictions of our models for simulated GW150914 were consistent with those obtained by LALInference for GW150914. The total estimation time was approximately six minutes, confirming that computational costs can be significantly reduced.  This is an easier task than using observational data, which includes real noise, and therefore does not allow for a rigorous comparison. However, regarding the total estimation time, it is feasible to achieve a similar level of efficiency when applied to real GW data, as this method does not require extensive waveform generation or likelihood evaluations.

We analyzed the differences in the attention regions between the effective spin estimation model and the chirp mass estimation model using attention maps.
As a result, we confirmed that our models focus on the ranges where the effects of each parameter are predominant.
These results suggest that the predictions of our models are based on physically meaningful information.

Currently, our results only indicate that the attention regions of our models are roughly consistent with the ranges where the effects of each parameter are predominant. By applying the fisher analysis conducted in \cite{fisher}, it may be possible to quantitatively demonstrate the degree to which they are consistent.

Furthermore, we demonstrated the potential of utilizing attention maps to quantify the impact of glitches on parameter estimation.
As a result, we found that as the models focus more on glitches, the parameter estimation results become more strongly biased.
This suggests that attention maps could potentially be used to distinguish between cases where the results produced by the machine learning model are reliable and cases where they are not.

It would be interesting to explore including data with glitches in the training set as a way to develop a model that is robust to glitches. In such a model, attention may not be directed toward the glitches. It would also be interesting to analyze the Attention Maps of our models after applying the glitch mitigation methods proposed in, e.g., \cite{glitch_mitigation}.

\mycomments{In this paper, we proposed that attention maps enhance the reliability of the gravitational wave parameter estimation against transient noises in the detector output called glitches. Leveraging the power of the attention maps requires an inspection of each map for each event. If the detection rate of compact binary star mergers is small, as is the case currently, we can manually inspect the attention maps of each event. When the detection rate becomes larger, as is expected to happen in near future as the detectors' sensitivities  improve, we have to develop an automatic way of assessing the attention maps.  One way to do that would be to check the consistency between the time of coalescence 
estimated from the matched filtering stage and the time when the attention map peaks. }

\mycomments{As for other technical details, we used Gaussian noise to estimate parameter errors. We leave use of real data as future work. 
This work is intended to show that attention maps enhance the reliability of the machine learning technique in gravitational wave data analysis. For this purpose, we used a spectrogram as it is easier to understand than a time series or its Fourier transform.
As a possible future direction, it would be interesting to use time series in addition to spectrogram to make a network learn phases of gravitational waves and glitches. Finally, we studied attention maps of the class token at the last output stage of the network. It would be interesting to use the attention roll-out that incorporates the information of the attention map of each layer \cite{2020Abnar}.}

\vspace{-15pt}
\section*{Acknowledgements}
\vspace{-10pt}
We would like to thank Nobuyuki Kanda, Masaki Uematsu, Izumi Kaku, Takumi Fujimori, Ryuki Kawamoto, and Hiroyuki Nakano for helpful comments.  We are grateful to Kota Tomita and Kyohei Takatani for their contribution in the early phase of this work. We would also like to thank Hirotaka Takahashi, Takashi Uchiyama, Yutaka Shikano, Shoichi Oshino and Yusuke Sakai for the valuable insights provided regarding the application of machine learning to gravitational wave data analysis. This work was supported by JSPS KAKENHI Grant No. 20H05639. We use PyTorch \cite{pytorch} for the implementation of our models. The plots are generated with matplotlib \cite{matplotlib}.

\appendix

\section{\mycommentsII{Deit}}
\label{sec:appendix_Deit}

\mycommentsII{In this work, we used the Data-efficient Image Transformers, specifically, the deit\_small\_distilled\_patch16\_224 model provided by the timm library, and we followed the default implementation for fine-tuning and inference. As described in Section II.B (Training and validation), we used DeiT models pre-trained on ImageNet. Therefore, we did not train a separate teacher network on the same tasks of mass and spin estimation. The basic architecture of the model is publicly available at the following site \cite{timm_deit}, and as shown there, in addition to the class token, the distillation token also contributes to the model’s predictions. }

\section{\mycomments{Evaluation of uncertainties}}
\label{sec:appendix}
\mycomments{
In this appendix, we give the idea behind and 
possible use cases of our method of evaluating parameter 
uncertainties
}

\mycomments{
We have used a simple frequentist approach. 
Let us explain it using a simple example.
Suppose our data $x$ consists of a noise $n$ and signal $\mu$ like $x=n+\mu$. 
We may estimate the value of $\mu$ from $x$. 
Now, if we have lots of statistically independent $x$'s ($x_1,x_2,\cdots,x_N$) for a fixed $\mu$, 
then we obtain $N$ point-estimates of $\mu$ from which we obtain an error estimate of our point estimate of $\mu$.  
To realize this, we need to have lots of noise realizations $\{ n_i \}_{i=1}^N$.
} 

\mycomments{
In the real-world application of our method, we propose that such noise realizations are obtained by injecting a simulated signal $h$ with fixed 
signal parameters $\hat p$ into time series starting at many different times, 
assuming that the statistical properties of the detector output time series are stable in time.
Indeed, this is what is done to estimate the false alarm rate of gravitational wave events in the literature. 
}

\mycomments{
The question then is how we find $\hat p$.  
A possible flow of the data analysis would be the following. 
The first process is the matched filtering, with which one searches for gravitational wave events 
and, if detected, which gives us point estimates of the astrophysical parameters such as the masses of the stars. 
One usually uses, e.g., Markov-Chain Monte-Carlo or nested sampling techniques to estimate the parameter errors, 
but that takes huge computational costs. One of the problems we are tackling in this paper is this computational cost problem. 
As such, to some extent, we can assume that we are given point estimates of parameters from the search pipeline 
(matched filtering stage) before we start. We then use that set of estimated parameters to generate a simulated waveform 
and inject it into sets of time series at many different time stamps (where no signal is known to be present). We obtain astrophysical parameters from each time series and error estimates from them.
} 


\bibliography{bibliography}

\end{document}